\documentclass[aps,prb,twocolumn,amsmath,amssymb,superscriptaddress,floatfix]{revtex4}

\usepackage{color}

\usepackage{graphicx}
\usepackage{subfigure}
\usepackage{bbold}
\usepackage{verbatim}
\usepackage{float}
\usepackage{enumerate}
\usepackage{amsfonts}
\usepackage{multirow}
\usepackage[colorlinks,bookmarks=false,citecolor=blue,linkcolor=red,urlcolor=blue]{hyperref}

\begin{document}

\title{
Spectral evidence of a ghostly transition: Theory of NMR 1/T$_1$ relaxation 
in a quantum spin nematic in applied magnetic field
}

\author{Andrew Smerald}

\affiliation{Institute of Physics, Ecole Polytechnique 
F{\'e}d{\'e}rale de Lausanne (EPFL), CH-1015 Lausanne, Switzerland}
\affiliation{Okinawa Institute of Science and Technology Graduate University, Onna-son, 
	Okinawa 904-0495, Japan}
\affiliation{H. H. Wills Physics Laboratory, University of Bristol, Tyndall Av., Bristol BS8 1TL, UK}

\author{Nic Shannon}

\affiliation{Okinawa Institute of Science and Technology Graduate University, Onna-son, 
	Okinawa 904-0495, Japan}
\affiliation{H. H. Wills Physics Laboratory, University of Bristol, Tyndall Av., Bristol BS8 1TL, UK}
\affiliation{Clarendon Laboratory, University of Oxford, Parks Rd. OX1 3PU, UK.}

\date{\today}

\begin{abstract}  
There is now strong theoretical evidence that a wide range of frustrated magnets 
should support quantum spin-nematic order in applied magnetic field.  
Nonetheless, the fact that spin-nematic order does not break time-reversal 
symmetry makes it very difficult to detect in experiment.
In this article, we continue the theme begun in 
[
Phys.~Rev.~B {\bf 88}, 184430 (2013)], 
of exploring how spin-nematic order reveals itself in the spectrum of spin excitations.
Building on an earlier analysis of inelastic neutron scattering
[
Phys.~Rev.~B {\bf 91}, 174402 (2015)],  
we show how the NMR $1/T_1$ relaxation rate could be used to identify
a spin-nematic state.   
We emphasise the characteristic, universal features of $1/T_1$, using a 
symmetry-based description of the spin-nematic order parameter and its fluctuations.
Turning to the specific case of spin-1/2 frustrated ferromagnets, 
we show that the signal from competing spin-wave excitations can be suppressed
through a judicious choice of nuclear site and field direction. 
As a worked example, we show how $^{31}$P NMR in the square-lattice frustrated 
ferromagnet BaCdVO(PO$_4$)$_2$ is sensitive to spin-nematic order.
\end{abstract}


\maketitle

\section{Introduction}
\label{sec:introduction}

In the quantum spin-nematic state a set of spin-quadrupole moments order, breaking spin-rotation 
symmetry but not time-reversal symmetry~\cite{blume69,chen71,andreev84,papanicolaou88,chubukov91,
barzykin91,lauchli06,tsunetsugu06,shannon06,sindzingre09,ueda09,shindou09,
shindou11,shindou13,smerald13,smerald15}.
%
While a number of theoretical models clearly demonstrate spin-nematic order, finding experimental 
evidence for such a state has proved difficult.
This is largely due to the absence of time-reversal symmetry breaking, which means that the order 
parameter does not couple to the usual probes of magnetism.
For example there are no Bragg peaks in elastic neutron scattering and no splitting of spectral 
lines in nuclear magnetic resonance (NMR) experiments.


Recently, we proposed that one way to detect the spin-nematic state would be via inelastic scattering 
of neutrons\cite{smerald-thesis,smerald15}.
This is based on the finding that excitations of the quadrupolar order parameter drive a small spin-dipole 
fluctuation\cite{smerald13}.
This can couple to dynamic probes of magnetism, and we predicted that a ghostly Goldstone mode 
excitation should be visible in neutron scattering experiments.


Here, we consider another dynamic probe of magnetic fluctuations, the NMR $1/T_1$ relaxation rate, 
and show how it could be used to identify spin-nematic order.
This could prove useful either as a complement to inelastic neutron scattering experiments, 
or in conditions where neutron scattering is not possible.
For example, in the presence of strong magnetic fields or when only small crystals are available.

\begin{figure}[t]
\centering
\includegraphics[width=0.45\textwidth]{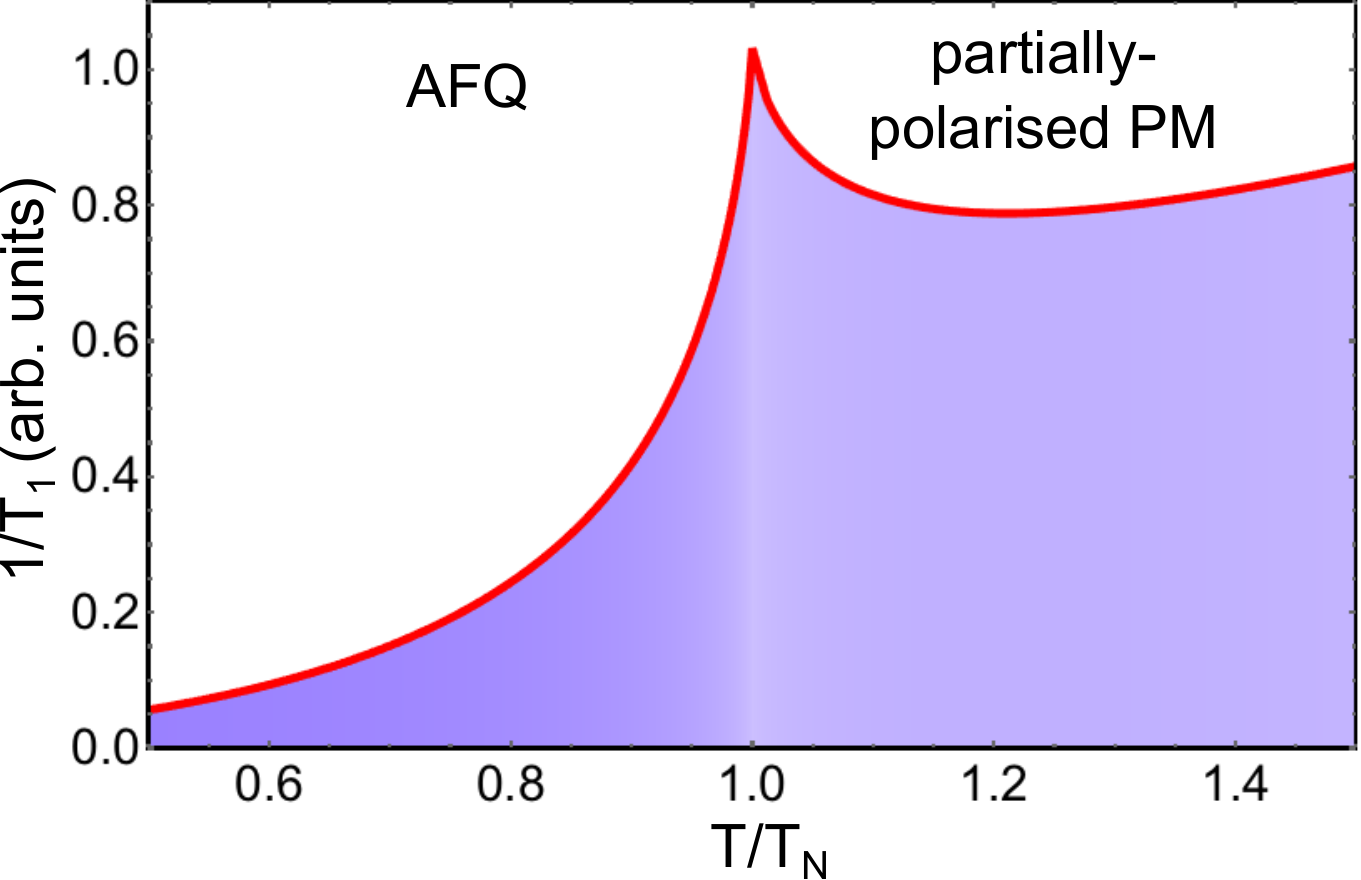}
\caption{\footnotesize{
Prediction for the NMR $1/T_1$ relaxation rate [see Eq.~\ref{eq:AFQT1}] close to a continuous 
phase transition from a partially-polarised paramagnet (PM) to an antiferroquadrupolar (AFQ) 
ordered state.
The contribution from long-wavelength fluctuations of the quadrupolar order parameter is shown.
To fully determine the NMR relaxation due to magnetic processes this should be combined with 
the contribution from gapped transverse excitations of the partially-polarised moment 
(see Section.~\ref{sec:transverseT1}).
There is a step-like jump in $1/T_1$ on crossing the critical temperature, and a sharp cusp 
at $T_{\sf Q}$, but no critical divergence.
In contrast, $1/T_1$ in a canted antiferromagnet shows divergent behaviour at the critical temperature (see Appendix~\ref{app:NMR-AFM}).
}}
\label{fig:AFQT1}
\end{figure}

The main result is shown in Fig.~\ref{fig:AFQT1}.
At a thermal phase transition between the spin-nematic state and the partially-polarised paramagnet 
there is a step-like increase in $1/T_1$ and a sharp cusp.
When combined with other considerations, this could provide strong evidence for the existence 
of spin-nematic order.


While the result shown in Fig.~\ref{fig:AFQT1} is a universal feature of spin-nematic order, we couch most of the discussion in this article in terms of frustrated, spin-1/2 ferromagnets in applied magnetic field.
The reason for this is that these appear to be one of the most promising places to search for the spin-nematic state experimentally\cite{sato09,zhitomirsky10,sato11,sato13,ueda13,smerald15,ueda15}.


The simplest theoretical model that supports spin-nematic order in spin-1/2 systems is,
\begin{align}
\mathcal{H}^{\sf S=1/2}_{\sf J_1-J_2}=&
J_1 \sum_{\langle ij \rangle_1} {\bf S}_i.{\bf S}_j 
+ J_2 \sum_{\langle ij \rangle_2} {\bf S}_i.{\bf S}_j 
- h \sum_i S_i^{\sf z},
\label{eq:HJ1J2}
\end{align}
where $\langle ij \rangle_1$ counts first-neighbour bonds and $\langle ij \rangle_2$ second-neighbour bonds. 
These bonds could, for example, live on a 1-dimensional (1D) chain\cite{chubukov91}, a 2-dimensional (2D) square lattice\cite{shannon06} or a 3-dimensional (3D) bcc lattice\cite{ueda13}.

\begin{figure}[t]
\centering
\includegraphics[width=0.45\textwidth]{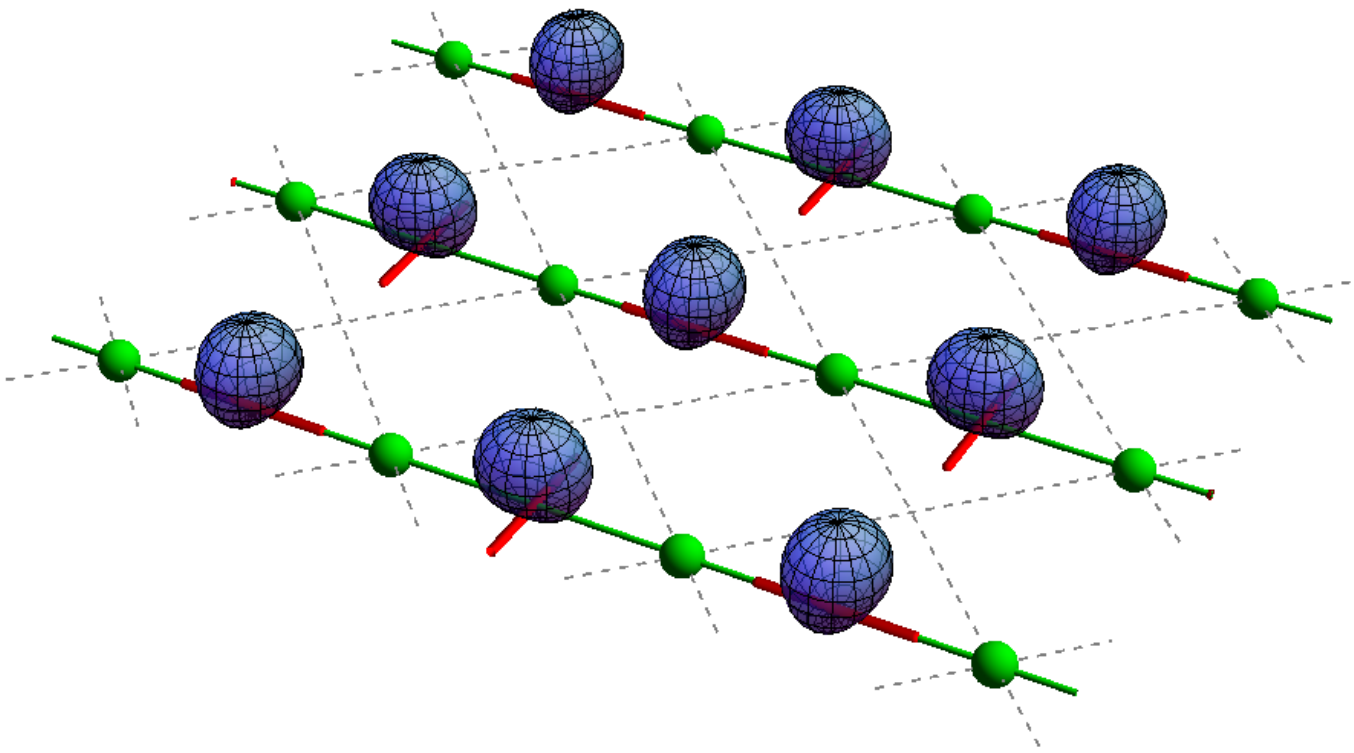}
\caption{\footnotesize{
Two-sublattice, bond-centered spin-nematic  state, 
of the type found to be the ground state of quasi-1D, spin-1/2 frustrated ferromagnets in magnetic field\cite{ueda09,sato13,starykh14,syromyatnikov12,zhitomirsky10}. 
For example, this has been proposed as the ground state of LiCuVO$_4$ close to saturation\cite{zhitomirsky10,svistov10,buttgen14}.
The probability distribution for the moment on each bond is shown as a blue surface.
The orthogonal `directors' associated with this type of nematic order are shown as red cylinders.
Green spheres represent magnetic atoms.
}}
\label{fig:SNchain}
\end{figure}

In the case of 1D, it has been shown that for ferromagnetic $J_1$, $\mathcal{H}^{\sf S=1/2}_{\sf J_1-J_2}$ has dominant quadrupolar correlations for a wide parameter range close to saturation\cite{chubukov87,chubukov91,hikihara08,sudan09,kecke07,kuzian07,vekua07,heidrich-meisner06,lu06}.
These can lead to distinctive 1D critical behaviour in the $1/T_1$ relaxation rate\cite{sato09,sato11}. 
The addition of a weak interchain coupling can then stabilise a long-range ordered quadrupole state\cite{ueda09,sato13,starykh14,syromyatnikov12,zhitomirsky10} (see Fig.~\ref{fig:SNchain}).
The material LiCuVO$_4$ is thought to be a realisation of this model, and may show spin-nematic order close to saturation\cite{svistov10,zhitomirsky10,buttgen14}.
However, high-field NMR measurements have not yet detected evidence for such a state, and have shown that, if it does exist, it is limited to a very narrow field range\cite{buttgen14}.
The material PbCuSO$_4$(OH)$_2$ (linarite) is also believed to be a realisation of this model, and may have a multipolar phase at high field\cite{schapers13,willenberg16}.

\begin{figure}[t]
\centering
\includegraphics[width=0.45\textwidth]{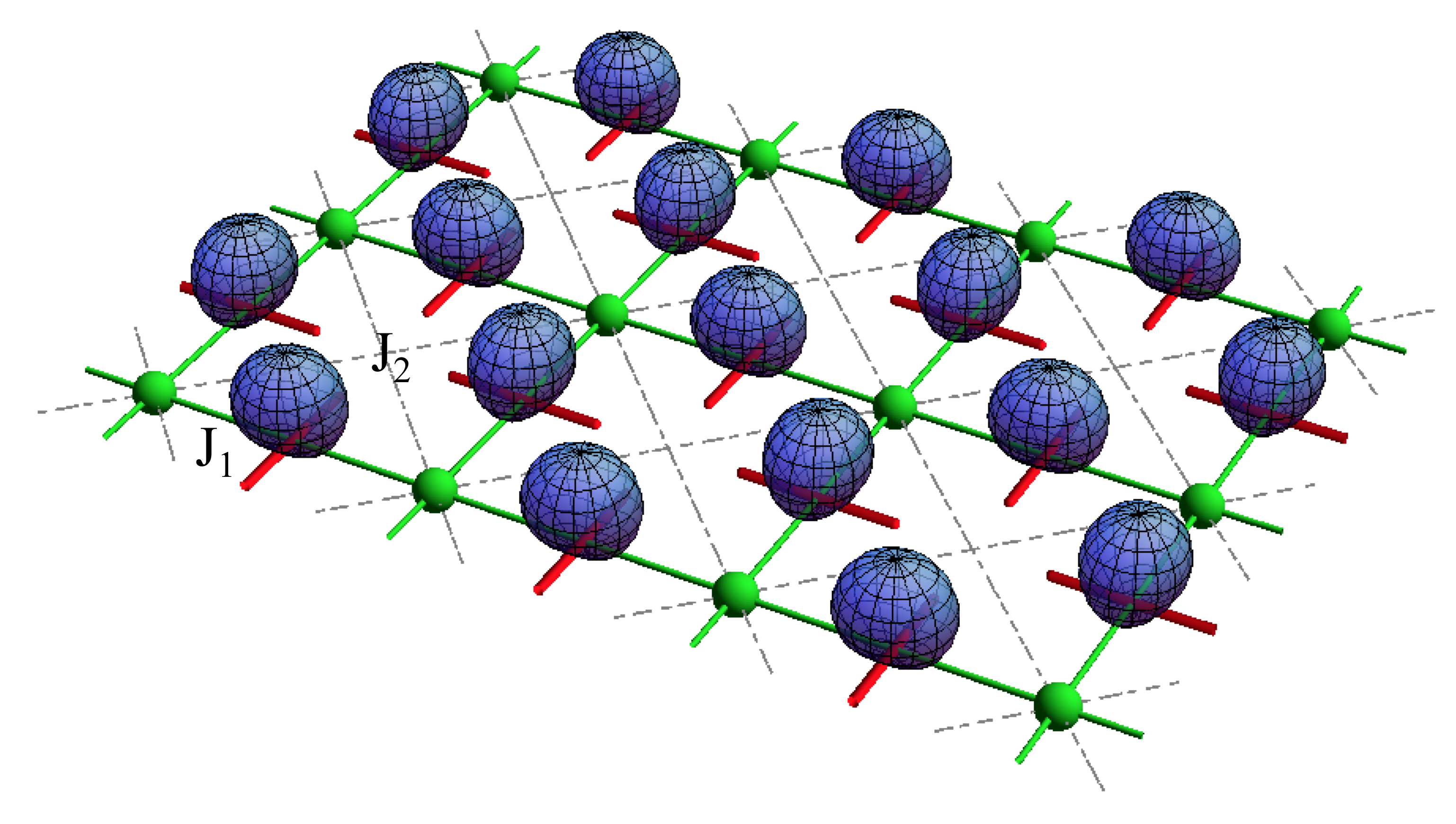}
\caption{\footnotesize{
Two-sublattice, bond-centered spin-nematic  state, 
of the type found to be the ground state of square-lattice, quasi-2D, spin-1/2 frustrated ferromagnets in magnetic field\cite{shannon06,shindou09,ueda07,shindou11,shindou13,smerald15}, reproduced from Ref.~[\onlinecite{smerald15}].
For example, this has been proposed as the ground state of BaCdVO(PO$_4$)$_2$ close to saturation\cite{smerald15}.
The probability distribution for the moment on each bond is shown as a blue surface.
The orthogonal `directors' associated with this type of nematic order are shown as red cylinders.
Green spheres represent magnetic atoms.
}}
\label{fig:SNsquare}
\end{figure}

In 2D a $J_1$-$J_2$ model on the square lattice has been shown to support spin-nematic order for a wide range of parameters close to saturation\cite{shannon06,shindou09,ueda07,sindzingre09,shindou11,shindou13,smerald15} (see Fig.~\ref{fig:SNsquare}).
The addition of weak interlayer coupling, can stabilise this phase at finite temperature\cite{ueda15}.
This model is approximately realised in a number of quasi-2D materials, including the vanadates Pb$_2$VO(PO$_4$)$_2$\cite{kaul04,kaul05,tsirlin09,skoulatos09,nath09}, SrZnVO(PO$_4$)$_2$\cite{tsirlin09,skoulatos09,bossoni11} and 
BaCdVO(PO$_4$)$_2$\cite{nath08,tsirlin09}.


The NMR $1/T_1$ relaxation measures fluctuations of the internal magnetic field at a nuclear site, and these arise predominantly from fluctuations of the neighbouring electronic spins.
In consequence, $1/T_1$ can be expressed in terms of the imaginary part of the dynamic susceptibility of the electronic spin system\cite{moriya56,beeman68,mila89}, with the most general form~\cite{smerald11},
\begin{align}
\frac{1}{T_1({\bf h}_{\sf ext})} 
 \hspace{-1mm} =  \hspace{-1.5mm} \lim_{\omega \rightarrow 0}
 \frac{\gamma_N^2}{2N}  k_{\sf B}T \hspace{-1mm} \sum_{{\bf q},\alpha,\beta} \hspace{-1mm}
    \mathcal{F}^{\alpha \beta} \hspace{-0.5mm}({\bf q},{\bf h}_{\sf ext})
\frac{ \Im m \left\{ \chi^{\alpha\beta}\hspace{-0.5mm}({\bf q},\omega) \right\} }{\hbar \omega},
\label{eq:1/T1}
\end{align}
where $\gamma_N$ is the nuclear gyromagnetic ratio, 
$\omega$ is the NMR frequency,
${\bf h}_{\sf ext}$ is the externally applied magnetic field, 
$\alpha$ and $\beta$ label spin components, 
and $ \mathcal{F}^{\alpha \beta}({\bf q},{\bf h}_{\sf ext})$ is a form factor 
describing the coupling between nuclear and electronic spins.


In the spin-nematic state, fluctuations of the quadrupolar order parameter drive spin-dipole fluctuations, and these couple to $1/T_1$ relaxation.
The characteristic behaviour of $1/T_1$ is shown in Fig.~\ref{fig:AFQT1} and we argue below that  measurement of such a feature could provide good evidence for spin-nematic order.
However, spin-dipole fluctuations associated with the quadrupolar Goldstone mode are not the only driver of $1/T_1$ relaxation.
In spin-1/2 frustrated ferromagnets, the spin-nematic state is expected to appear close to saturation, and therefore coexists with a partially-polarised moment.
Transverse fluctuations of this partially-polarised moment can have a small gap, and therefore contribute significantly to $1/T_1$ at temperatures comparable to the spin-nematic ordering temperature.
The danger is that these transverse fluctuations could swamp the contribution of the quadrupolar Goldstone mode to relaxation.
Thus it would be useful to have a way of suppressing this effect, and we show how the $1/T_1$ form factor can be used to ``filter out'' transverse fluctations of the polarised moment.


The remainder of this paper is organised as follows.
In Section~\ref{sec:SNintro} the spin-nematic state in frustrated spin-1/2 ferromagnets is reviewed, with particular attention to the dispersion of magnetic excitations.
Section~\ref{sec:QuadT1} contains the main result, the $1/T_1$ response due to fluctuations of the quadrupolar order parameter.
Section~\ref{sec:transverseT1} considers the main competing contribution to $1/T_1$, transverse fluctuations of the partially-polarised moment, and shows how these can be suppressed by the NMR form factor.
Section~\ref{sec:formfactor} shows a worked example of form factor suppression, which is relevant to P NMR in the material BaCdVO(PO$_4$)$_2$.
Section~\ref{sec:lowT} studies the low-T power-law behaviour of $1/T_1$ in a spin-nematic state.
Finally, in Section.~\ref{sec:conclusion} we discuss the likelihood of observing the $1/T_1$ features shown in Fig.~\ref{fig:AFQT1}, and how characteristic these are of the spin-nematic state. 

\section{Magnetic excitations of the spin-nematic state}
\label{sec:SNintro}

In order to determine the NMR $1/T_1$ relaxation rate, it is first necessary to understand the nature of the spin-nematic state.
The aim of this section is to provide an overview of what is currently known, in particular about the magnetic excitations, as this is what is probed by $1/T_1$ relaxation.
We focus on antiferroquadrupolar (AFQ) order in spin-1/2 frustrated ferromagnets close to saturation, since this appears to be the context in which spin-nematic order is most likely to be realised experimentally (see discussion below $\mathcal{H}^{\sf S=1/2}_{\sf J_1-J_2}$ [Eq.~\ref{eq:HJ1J2}]).


A typical Hamiltonian in which theory shows the existence of spin-nematic order is given by $\mathcal{H}^{\sf S=1/2}_{\sf J_1-J_2}$ [Eq.~\ref{eq:HJ1J2}].
An AFQ state is found to exist for ferromagnetic $J_1$, competing antiferromagnetic $J_2$ and magnetic fields close to saturation\cite{chubukov91,shannon06,ueda13}.

\begin{figure}[t]
\centering
\includegraphics[width=0.4\textwidth]{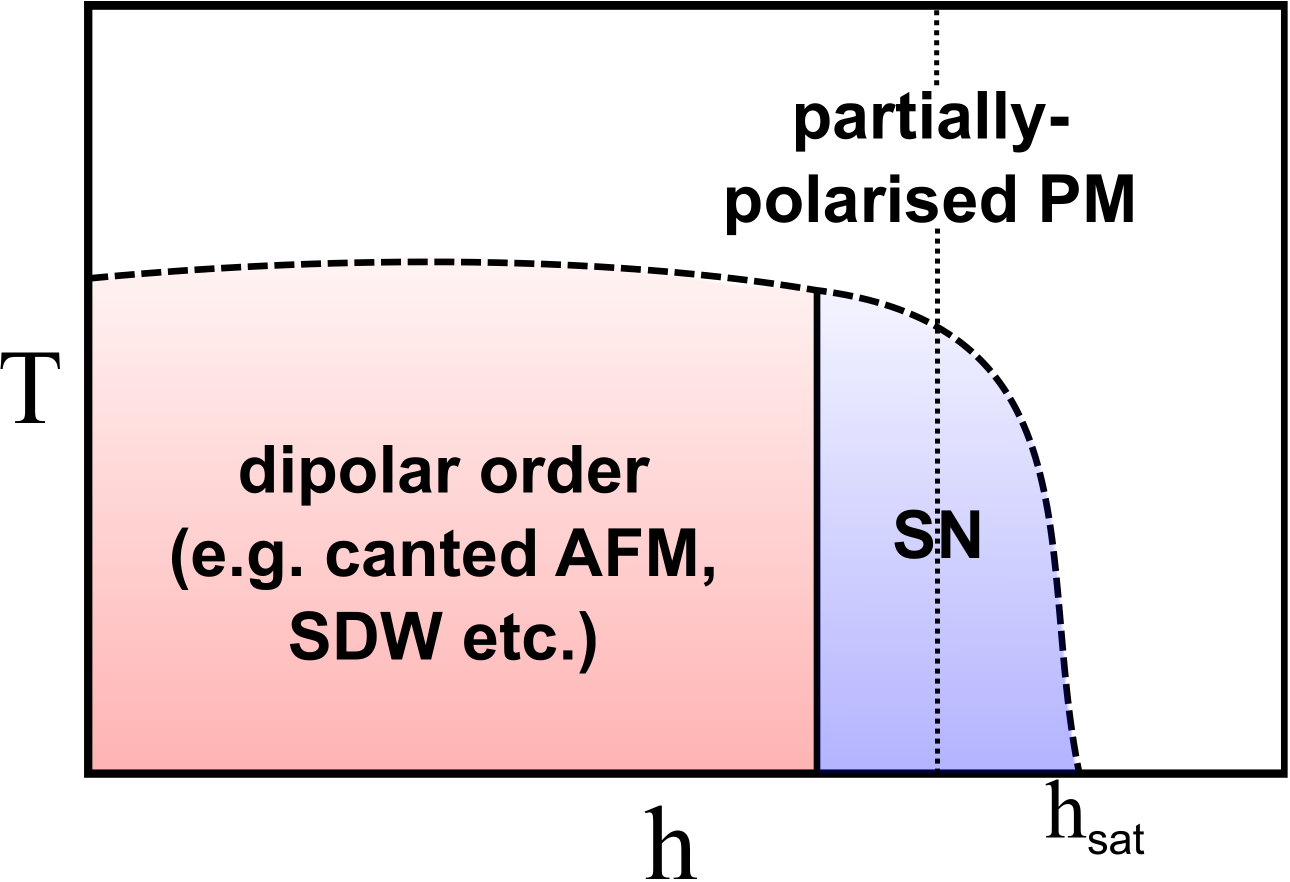}
\caption{\footnotesize{
Schematic phase diagram for spin-1/2 frustrated ferromagnets in magnetic field.
A spin nematic (SN) is typically found in theoretical models just below the saturation field, $h_{\sf sat}$\cite{ueda09,sato13,starykh14,syromyatnikov12,shannon06,shindou09,ueda07,shindou11,shindou13,smerald15}.
At lower fields a (set of) antiferromagnetically ordered state(s) is stabilised, for example a canted antiferromagnet or a spin-density wave (SDW) state.
Typically one would expect transitions from the partially-polarised paramagnet (PM) to the SN to be second order, while the transition between the SN and canted AFM is likely to be first order.
The dotted black line shows the path of a $1/T_1$ experiment.
}}
\label{fig:phasediag}
\end{figure}

A schematic phase diagram is shown in Fig.~\ref{fig:phasediag}.
At zero magnetic field and low temperature there is a magnetically ordered phase.
For the example of $\mathcal{H}^{\sf S=1/2}_{\sf J_1-J_2}$ [Eq.~\ref{eq:HJ1J2}]  on the square lattice, this is the columnar antiferromagnet with ordering vector ${\bf q}=(\pi,0)$ or ${\bf q}=(0,\pi)$.
Increasing the magnetic field causes the spins to cant towards the field direction, and in the presence of small anisotropies there can be spin-flop transitions between different magnetically ordered states.
However, the details of the low-field phases are not pertinent to the current discussion.
The important feature is that close to saturation there is a spin-nematic state, which is typically found to be a two-sublattice, bond-centred AFQ phase (see Fig.~\ref{fig:SNsquare}).

\begin{figure}[t]
\centering
\includegraphics[width=0.49\textwidth]{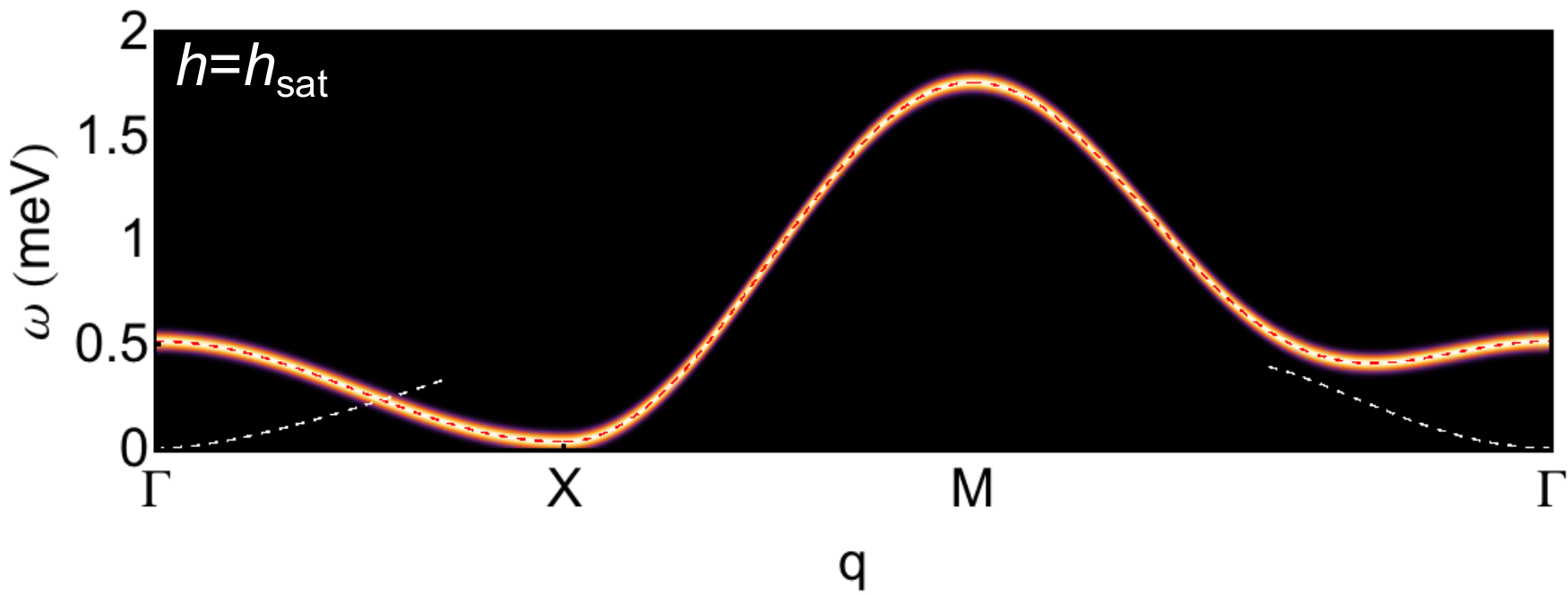}
\caption{\footnotesize{
Magnetic dispersion in the $J_1$-$J_2$ model [Eq.~\ref{eq:HJ1J2}] on the square lattice at the saturation field, $h_{\sf sat}$, reproduced from Ref.~[\onlinecite{smerald15}].
The parameters used are \mbox{$J_1=-3.6$K} and \mbox{$J_2=3.2$K}, and the path through the Brillouin zone is ${\bf q}_{\sf \Gamma}=(0,0)$ $\to$ ${\bf q}_{\sf X}=(\pi,0)$ $\to$ ${\bf q}_{\sf M}=(\pi,\pi)$ $\to$ ${\bf q}_{\sf \Gamma}$.
A dashed white line shows the gapless dispersion of 2-magnon bound states, while a dashed red line shows the gapped dispersion of 1-magnon transverse excitations of the polarised moment.
The colour scheme shows the imaginary part of the dynamic spin susceptibility perpendicular to the magnetic field.
}}
\label{fig:chiperp}
\end{figure}

The best way to understand the spin nematic state is to first consider the saturated paramagnet\cite{shannon06,ueda09,ueda15,smerald15}.
In the saturated paramagnet all excitations are gapped.
As the field is lowered towards the saturation value, there is a gap that closes, and the associated excitation condenses, forming an ordered state.
In spin-1/2 frustrated ferromagnets it can happen that bound-magnon pairs condense, and this results in the formation of a quadrupolar state (see Fig.~\ref{fig:chiperp}).


The order parameter of the spin-nematic state is a rank-2, symmetric, traceless tensor\cite{andreev84},
\begin{align}
Q_{ij}^{\alpha\beta} = S_i^\alpha S_j^\beta + S_i^\beta S_j^\alpha -\frac{2}{3}\delta^{\alpha\beta} {\bf S}_i \cdot {\bf S}_j,
\label{eq:OP}
\end{align}
with $\alpha,\beta = {\sf x,y,z}$. 
The creation of a bound-magnon pair in the saturated state can be described by\cite{shannon06},
\begin{align}
S^-_i S^-_j | \mathrm{sat} \rangle
&= \left[ 
\left(S^{\sf x}_i S^{\sf x}_j -S^{\sf y}_i S^{\sf y}_j\right)  
- i  \left(S^{\sf x}_i S^{\sf y}_j +S^{\sf y}_i S^{\sf x}_j\right)  
\right]  | \mathrm{sat} \rangle \nonumber \\
&= \left[ 
Q^{\sf x^2-y^2}_{ij}
- i  Q^{\sf xy}_{ij}
\right]  | \mathrm{sat} \rangle,
\end{align}
where $| \mathrm{sat} \rangle$ is the fully saturated state and,
\begin{align}
Q^{\sf x^2-y^2}_{ij} = (Q^{\sf xx}_{ij} - Q^{\sf yy}_{ij})/2.
\label{eq:Qxsqysq}
\end{align}
Thus it can be seen that condensation of bound-magnon pairs leads to non-zero components in the order-parameter tensor $Q_{ij}^{\alpha\beta}$ [Eq.~\ref{eq:OP}].
This order parameter lives on the bonds, and is associated with triplet pairing of the underlying spin-1/2 moments.


In the saturated paramagnet it is possible to exactly calculate the excitation spectrum of $\mathcal{H}^{\sf S=1/2}_{\sf J_1-J_2}$ [Eq.~\ref{eq:HJ1J2}]\cite{smerald15}.
In Fig.~\ref{fig:chiperp}, which is reproduced from Ref.~[\onlinecite{smerald15}], the dispersion is shown for a square-lattice model at the saturation field, $h_{\sf sat}$.
It can be seen that there is gapless mode at ${\bf q}_{\sf \Gamma}=(0,0)$, and this is associated with bound-magnon pairs.
For $h<h_{\sf sat}$ this mode becomes the Goldstone mode of the spin-nematic state, and is primarily associated with rotation of the quadrupole moment in the plane perpendicular to the field.
However, the dynamics of this quadrupole rotation mixes a small spin-dipole character into the wavefunction\cite{smerald13,smerald15}.


In Fig.~\ref{fig:chiperp} one can also observe a set of 1-magnon excitations, which describe transverse fluctuations of the polarised moment.
At $h=h_{\sf sat}$ these excitations are gapped for all ${\bf q}$, but the gap is very small at ${\bf q}_{\sf X} = (\pi,0)$.
The condensation of magnons at this wavevector is a competing instability of the saturated paramagnet, but is preceded by the condensation of bound-magnon pairs.
If it were the leading instability, then a canted antiferromagnet would form, with ordering vector ${\bf q}_{\sf X} = (\pi,0)$.
In the spin-nematic state this band of 1-magnon excitations remains gapped, and is very little changed with respect to the saturated paramagnet. 


One way to understand the excitation spectrum within the spin-nematic state in 2D is via a lattice gauge theory\cite{shindou09,shindou11,shindou13}.
This takes into account the highly entangled nature of the bond-nematic state, and shows that it is analagous to the resonating valence-bond state, only with singlets replaced by triplets.
This can in principle be used to describe all the different excitations of the spin-nematic state.
However, it has not been fully developed in the presence of magnetic field.


Here we follow Ref.~[\onlinecite{smerald13,smerald15,smerald-thesis}] and instead consider a continuum theory of the quadrupolar order parameter.
This can be used to describe the Goldstone mode excitations, and has the advantage of bringing the universal properties to the fore.
We complement this by separately considering the 1-magnon excitations.
These remain gapped throughout the spin-nematic state, and can be well approximated by considering the dispersion at saturation. 

\section{1/T$_1$ relaxation due to fluctuations of spin-nematic order}
\label{sec:QuadT1}

First we consider the NMR $1/T_1$ relaxation due to dipolar fluctuations associated with the spin-nematic Goldstone mode (see Fig.~\ref{fig:chiperp}).
We study in particular the critical region, close to a second-order thermal phase transition between the spin-nematic and partially-polarised paramagnetic states (see dotted line in Fig.~\ref{fig:phasediag}).
We find that while there is no divergence of $1/T_1$ at the critical point, it does display a step-like increase and a sharp cusp, as shown in Fig.~\ref{fig:AFQT1}.


In conventional antiferromagnets the order parameter is dipolar in nature.
At the critical point the correlation length diverges, and there is an associated critical slowing down of spin fluctuations.
This leads to a divergence in $1/T_1$\cite{moriya62,borsa-book,ziolo88,engelsberg80}, and the theory underpinning this is summarised in Appendix~\ref{app:NMR-AFM}.


For an AFQ in applied magnetic field the order parameter is quadrupolar in nature and is confined to the plane perpendicular to the field direction.
For field applied in the {\sf z} direction it is given by,
\begin{align}
 {\bf Q}_\perp({\bf r}) =
 \left(
\begin{array}{c}
Q_{\sf A}^{\sf x^2-y^2}({\bf r}) - Q_{\sf B}^{\sf x^2-y^2}({\bf r}) \\
Q_{\sf A}^{\sf xy}({\bf r}) - Q_{\sf B}^{\sf xy}({\bf r})
\end{array}
\right),
\label{eq:QperpOP}
 \end{align}
where $Q_\mu^{\sf x^2-y^2}({\bf r})$ is the spatial average of the bond-based operator $Q_{ij}^{\sf x^2-y^2}$ [Eq.~\ref{eq:Qxsqysq}] over bonds in the $\mu \in \{ {\sf A}, {\sf B}\}$ sublattice, and similarly for $Q_\mu^{\sf xy}({\bf r})$ [Eq.~\ref{eq:OP}].
The average is taken over a region of space much larger than the lattice constant and at least comparable to the correlation length, $\xi$.


Symmetry constrains the associated Landau theory to be,
\begin{align}
\mathcal{H}^{\sf Lan}_{\sf AFQ} = 
 \frac{\alpha\tilde{t}}{2} {\bf Q}_\perp^2  
 + \frac{u}{4} {\bf Q}_\perp^4 + \dots ,
\label{eq:LandauTheory}
\end{align}
where $\alpha > 0$ and and $\tilde{t} = (T-T_{\sf N})/T_{\sf N}$ is the reduced temperature.
A second-order phase transition at $\tilde{t} = 0$ between the AFQ and the partially-polarised 
paramagnet is thus allowed by symmetry, and this is what we consider here.
While it is possible that the transition can be driven to be weakly first order by fluctuations, 
this will not qualitatively change the conclusions.


In order to calculate the dynamic spin susceptibility we first consider the static Ginzburg-Landau functional, and then add dynamics in a phenomenological manner.
The Ginzburg-Landau functional is given by\cite{barzykin91,shannon10},
\begin{align}
\mathcal{H}^{\sf GL}_{\sf AFQ} &\approx \hspace{-1mm} \int d^3r \left[ \frac{\alpha\tilde{t}}{2} {\bf Q}_\perp^2  
+ \frac{K}{2} (\nabla {\bf Q}_\perp)^2 + \frac{u}{4} {\bf Q}_\perp^4  +\frac{1}{2\chi_{\sf z}^{\sf Q}}  (l^{\sf z})^2
\right]
\label{eq:GL-AFQ}
\end{align}
where $K$ is a generalised elastic constant, 
$\chi_{\sf z}^{\sf Q}$ is the quadrupolar susceptibility, $l^{\sf z}$ is a canting field describing spin-dipole fluctuations parallel to the applied magnetic field\cite{smerald13,smerald-thesis} and an irrelevant coupling $hl^{\sf z}$ has been ignored.
The coupling $u$ is now assumed to be positive, consistent with a continuous phase transition.
For simplicity we have assumed that the AFQ state is isotropic in 3D.
While this is clearly not the case for quasi-1D and quasi-2D materials, it is a simple exercise to introduce spatial anisotropies into the model and we have checked that it does not make a qualitative difference to predictions for $1/T_1$. 


Close to the transition the correlation length scales as $\xi \propto \tilde{t}^{-\nu}$, and in mean-field theory $\nu=1/2$.
However, since the order parameter does not produce an internal magnetic field, this critical divergence of the correlation length does not drive a divergence of the $1/T_1$ relaxation rate.
Instead it is the small dipolar fluctuations, $l^{\sf z}$, dynamically generated by rotations of the quadrupoles, that couple to $1/T_1$.
These are not critical, and are completely suppressed at the ordering vector\cite{smerald13}.


It is useful to separate longitudinal and transverse fluctuations, and this can be acheived by writing, 
\begin{align}
Q_\mu^{\sf x^2-y^2}=\eta [(n^{\sf x}_\mu)^2-(n^{\sf y}_\mu)^2], \quad
Q_\mu^{\sf xy}=2\eta n^{\sf x}_\mu n^{\sf y}_\mu,
 \end{align}
where ${\bf n}_\mu({\bf r})=(n_\mu^{\sf x},n_\mu^{\sf y},0)$ is a unit vector aligned with the nematic directors (see Fig.~\ref{fig:SNchain}) and 
\mbox{${\bf n}_{\sf A}({\bf r}) \cdot {\bf n}_{\sf B}({\bf r})=0$}.
This results in,
%
\begin{align}
\mathcal{H}_{\sf AFQ}^{\sf GL}
   &\approx
   \int d^3r \left[ \frac{\alpha\tilde{t}}{2}\eta^2  
   + \frac{K}{2} (\nabla \eta)^2 + \frac{u}{4}\eta^4 \right. 
   \nonumber \\
   & \left. +  \frac{\rho_{\sf d}(\eta)}{2}  \left( (\nabla{\bf n}_{\sf A})^2 +(\nabla{\bf n}_{\sf B})^2 \right)  
   + \frac{1}{2\chi_{\sf z}^{\sf Q}} (l^{\sf z})^2
    \right],
\label{eq:GL-AFQ-eta}
\end{align}
with the director stiffness, $\rho_{\sf d}(\eta)=\eta^2K$.


Dynamics can be added to the static Ginzburg-Landau model [Eq.~\ref{eq:GL-AFQ-eta}] according to the theory of dynamic critical phenomena in stochastic models\cite{chaikin-book,ma-book,halperin69a,halperin69b,hohenberg77} (see Appendix~\ref{app:NMR-AFM} for a similar treatment of the canted antiferromagnet).
The longitudinal fluctuations of the order parameter are not conserved and obey a purely dissipative equation of motion,
\begin{align}
 \partial_t \delta\eta({\bf r},t) &\approx -\Gamma K (\xi^{-2} -\nabla^2 ) \delta\eta({\bf r},t) +\zeta_{\eta}({\bf r},t),
 \end{align}
where $\delta \eta({\bf r},t)  = \eta({\bf r},t)  - \langle \eta \rangle$ and the mean-field approximation $u\rightarrow 0$ and $\nu=1/2$ has been assumed.
The phenomenological parameter $\Gamma^{-1}$, sets the rate of damping and $\zeta_\eta$ is a white noise term with correlation function,
\begin{align}
\langle \zeta_\eta({\bf r},t) \zeta_\eta({\bf r}^\prime,t^\prime) \rangle 
&= 2 \Gamma k_{\sf B}T \delta({\bf r}-{\bf r}^\prime) \delta(t-t^\prime),
\end{align}
which describes interactions between the long wavelength fluctuations of interest and short-wavelength excitations.
It follows that the correlation time is given by \mbox{$\tau_\eta({\bf q})\approx\Gamma^{-1}\chi_{\bf q}$}, 
with \mbox{$\chi_{\bf q}=K^{-1}(\xi^{-2}+ {\bf q}^2)^{-1}$}.
Approaching $T_{\sf Q}$,  $\tau_\eta(0)$ diverges as $\tau_\eta(0) \propto \xi^z$ with mean-field
dynamical exponent $z=2$.


In order to calculate the critical behaviour of $1/T_1$ in a canted antiferromagnet it is sufficient to model the dynamics of the longitudinal order parameter fluctuations.
This is because these fluctuations couple to the nuclear-spin lattice, leading to a critical divergence in $1/T_1$.
However, the AFQ order parameter is time-reversal invariant, and therefore its longitudinal fluctuations do not couple to NMR relaxation.
In consequence it is necessary to consider the long-wavelength ``flavour-wave'' excitations of the AFQ ordered moment.
These are the generalisation of spin waves in the antiferromagnet and are described by the fields ${\bf n}_\mu$ and $l^{\sf z}$ [\onlinecite{smerald13}].


The ${\bf n}_\mu$ fields can be parameterised as\cite{smerald13,smerald-thesis},
\begin{align}
{\bf n}_{\sf A} &= (\sqrt{1-\phi^2},\phi,0) \nonumber \\
{\bf n}_{\sf B} &= (-\phi,\sqrt{1-\phi^2},0),
\end{align}
and we make the linear approximation $\sqrt{1-\phi^2} \approx 1$.
The coupled equations of motion are,
\begin{align}
\partial_t \phi({\bf r},t) &\approx  \frac{l^{\sf z}({\bf r},t)}{\hbar\chi_{\sf z}^{\sf Q}} 
   + \gamma \rho_{\sf d}\nabla^2 \phi({\bf r},t) +\zeta_{\phi}({\bf r},t) 
   \nonumber \\
\partial_t l^{\sf z}({\bf r},t) &\approx  \frac{\rho_{\sf d}}{\hbar} \nabla^2\phi({\bf r},t)  
   + \frac{\lambda}{\chi_{\sf z}^{\sf Q}} \nabla^2 l^{\sf z}({\bf r},t) +\zeta_{l}({\bf r},t).
\label{eq:AFQ-EOM}
\end{align}
If only the first term in each equation is taken into account, they describe the coherent motion of flavour waves and are equivalent to the non-linear sigma model description of the AFQ dynamics\cite{smerald13,smerald-thesis}.
The phenomenological parameters $\gamma$ and $\lambda$ set the relaxation timescales for the fields $\phi$ and $l^{\sf z}$.
The auxiliary fields $\zeta_\phi$ and $\zeta_l$ parameterise the damping of long-wavelength fluctuations by interaction with short-wavelength excitations, and obey the correlation functions, 
\begin{align}
\langle \zeta_\phi({\bf r},t) \zeta_\phi({\bf r}^\prime,t^\prime) \rangle 
   &=2 \gamma k_{\sf B}T \delta({\bf r}-{\bf r}^\prime) \delta(t-t^\prime) 
   \nonumber \\
\langle \zeta_{\sf l}({\bf r},t) \zeta_{\sf l}({\bf r}^\prime,t^\prime) \rangle 
   &=2 \lambda k_{\sf B}T \nabla^2 \delta({\bf r}-{\bf r}^\prime) \delta(t-t^\prime).
\end{align}
Fluctuations of the total magnetisation commute with $\mathcal{H}_{\sf AFQ}^{\sf GL}$~[Eq.~\ref{eq:GL-AFQ-eta}],
and therefore the ${\bf q}=0$ component of $l^{\sf z}$ is conserved.


Solving the linearised equations of motion, Eq.~\ref{eq:AFQ-EOM}, results in a dispersion,
\begin{align}
\omega_{\bf q}^\pm \approx \pm v q -\frac{iD q^2}{2} , \ 
v=\sqrt{\frac{\rho_{\sf d}}{\chi_{\sf z}^{\sf Q}}}, \ 
D= \frac{\lambda}{\chi_{\sf z}^{\sf Q}}+\rho_{\sf d} \gamma,
\end{align}
where the first term in $\omega_{\bf q}^\pm$ describes the coherent motion of the quadrupoles 
and the second term is dissipative.


Fluctuations of the spin-dipole moments can be parametrised by,
\begin{align}
\delta S^{\sf z}({\bf r},t) =  l^{\sf z}({\bf r},t) +  [\langle\eta \rangle+ \delta\eta({\bf r},t)] l^{\sf z}({\bf r},t) + \dots.
\end{align}
As far as $1/T_1$ relaxation is concerned the first term gives a negligible contribution.
This is because $1/T_1$ probes the spin susceptibility at $\omega\rightarrow 0$ [cf. Eq.~\ref{eq:1/T1}], 
and at low energies the $l^{\sf z}$ term is suppressed by the fact that \mbox{$l^{\sf z}({\bf q}=0,t)$} is a conserved 
quantity.   


\begin{figure}[t]
\centering
\includegraphics[width=0.3\textwidth]{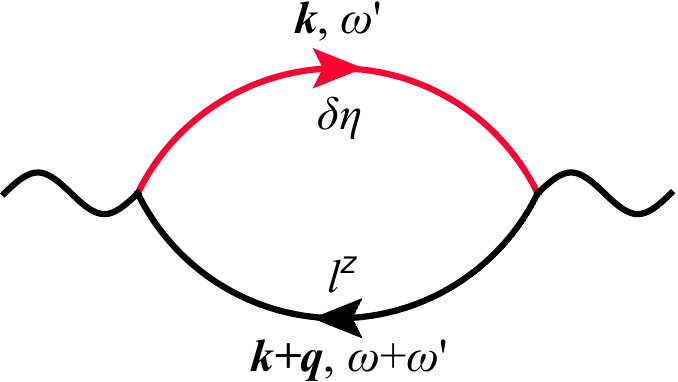}
\caption{\footnotesize{
Feynman diagram used to calculate the dynamic spin suscpeptibility [Eq.~\ref{eq:AFQchi}] close to the critical point of the antiferroquadrupolar (AFQ) spin-nematic state.
The upper line (red) is associated with longitudinal fluctuations of the order parameter, ${\bf Q}_\perp$ [Eq.~\ref{eq:QperpOP}], while the lower line (black) with dipolar fluctuations that are driven by the dynamics of ${\bf Q}_\perp$.
}}
\label{fig:eta-l-diagram}
\end{figure}


In order to determine $1/T_1$ [Eq.~\ref{eq:1/T1}] it is first necessary to calculate the imaginary part of the dynamical spin susceptibility.
At leading order this is given by the diagram shown in Fig.~\ref{fig:eta-l-diagram}, resulting in,
\begin{align}
&\frac{k_{\sf B}T}{\hbar\omega}  \Im m \{ \chi^{zz}_{\eta l}({\bf q},\omega) \}
 \approx 
 2 (g_{\sf l}\mu_{\sf B})^2 
\frac{\hbar(k_{\sf B}T)^2}{(k_{\sf B}T_{\sf N})^3} \int \frac{d^3k}{(2\pi)^3} \int \frac{d\omega^\prime}{2\pi}  
 \nonumber \\
&\frac{\Gamma}{\Gamma^2\chi_{\bf k}^{-2}+(\omega^\prime)^2}
 \frac{\gamma \rho_{\sf d}^2 ({\bf k}+{\bf q})^4
+\lambda (\omega+\omega^\prime)^2({\bf k}+{\bf q})^2}
{|(\omega+\omega^\prime-\omega^+_{{\bf k}+{\bf q}}) (\omega+\omega^\prime-\omega^-_{{\bf k}+{\bf q}})|^2}.
\label{eq:AFQchi}
\end{align}
Physically, this equation describes a longitudinal fluctuation of the quadrupole order parameter, 
which then cants, resulting in a spin-dipole fluctuation.
In order for this process to occur, it is necessary that $\tau_\eta \gg \tau_l$, where the correlation time 
for the $l^{\sf z}$ field is given by \mbox{$\tau_l({\bf q})\approx 2\chi_{\sf z}^{\sf Q}/\lambda {\bf q}^2$} 
in the paramagnet.
Close to the critical point this is equivalent to requiring $\chi_{\sf z}^{\sf Q} \Gamma K /\lambda \ll 1$.


Substituting Eq.~\ref{eq:AFQchi} into Eq.~\ref{eq:1/T1} results in a relaxation rate,
\begin{align}
&\frac{1}{T_1^{\sf AFQ}}
 \approx 
\gamma_{\sf N}^2   \frac{(g_{\sf l}\mu_{\sf B})^2}{8\pi^5}  \mathcal{F}_{\sf Q}  \hbar 
\frac{(k_{\sf B}T)^2}{(k_{\sf B}T_{\sf N})^3}
\int  dk_1  dk_2 d\omega  
\frac{\Gamma k_1^2}{\Gamma^2\chi_{{\bf k}_1}^{-2}+\omega^2}  \nonumber \\
& \frac{k_2^2\left(\gamma \rho_{\sf d}^2  k_2^4
+\lambda \omega^2 k_2^2\right)}
{\left(\left(\omega-v k_2\right)^2+\frac{D^2 k_2^4}{4}\right)
\left(\left(\omega+v k_2\right)^2+\frac{D^2 k_2^4}{4}\right)},
 \label{eq:AFQT1}
\end{align}
where the form factor, $\mathcal{F}_{\sf Q}=\mathcal{F}^{\alpha\beta}({\bf q}_{\sf nem},{\sf h}_{\sf ext})$ [Eq.~\ref{eq:1/T1}], is assumed constant at the nematic ordering vector, ${\bf q}_{\sf nem}$.
%


A numerical integration of this expression for an isotropic, 3D AFQ state results in the temperature 
dependence shown in Fig.~\ref{fig:AFQT1}.
It can be seen that there is a step-like increase in $1/T_1^{\sf AFQ}$ [Eq.~\ref{eq:AFQT1}] on crossing the critical point at $T_{\sf Q}$.
This is accompanied by a sharp cusp, but there is no critical divergence.


\section{1/T$_1$ relaxation due to transverse fluctuations}
\label{sec:transverseT1}

Fluctuations of the AFQ Goldstone mode are not the only excitations contributing to $1/T_1$, and we here consider relaxation due to transverse fluctuations of the partially-polarised moment.
From the perspective of detecting the existence of spin-nematic order, the danger is that these could swamp the signal from the AFQ Goldstone mode excitations.
In order to reduce this danger, we show how their contribution to $1/T_1$ could be suppressed by a careful choice of nuclear site and/or magnetic field direction.


In saturated, spin-1/2, frustrated ferromagnets, there is typically more than one competing instability on lowering the field\cite{shannon06,ueda13,smerald15,ueda15}.
The ``losing'' instabilities remain gapped at the saturation field, $h_{\sf sat}$, but this gap can be small.
An example of this is shown in Fig.~\ref{fig:chiperp}, where for $\mathcal{H}^{\sf S=1/2}_{\sf J_1-J_2}$ [Eq.~\ref{eq:HJ1J2}] on the square lattice there are 1-magnon modes at ${\bf q}=(\pi,0)$ and ${\bf q}=(0,\pi)$ with very small gaps.
At temperatures large compared to these gaps, the contribution of these excitations to $1/T_1$ can be significant.


In the saturated state the spectrum of transverse fluctuations of the polarised moment can be calculated exactly\cite{smerald15}.
We make the assumption that this band of excitations is not modified significantly at the zero-temperature phase transition between the saturated state and the spin-nematic state, and similarly at the thermal phase transition between the AFQ and the partially-polarised paramagnet.
This assumption seems reasonable, since the critical physics is associated with the 2-magnon band, and the 1-magnon excitations remain gapped in the AFQ state.


Following Ref.~[\onlinecite{smerald11}], the relaxation rate due to the 1-magnon transverse excitations can be calculated as,
\begin{align}
\frac{1}{T_1^{\sf 1-mag}}=
\mathcal{F}_{\sf 1-mag}
\frac{ (g\mu_B)^2  \hbar  \gamma_N^2   V_{\sf cell}^2 \Delta_1^5}{2\pi^3\bar{v}_s^6} 
\Phi \left( \frac{k_B T}{\Delta_1}  \right),
\end{align}
where $\Delta_1$ is the gap to 1-magnon excitations, $\bar{v}_s$ denotes the spatially averaged spin-wave velocities close to this gapped mode, the form factor $\mathcal{F}_{\sf 1-mag}$ is assumed to be finite and,
\begin{align} 
\Phi (x) =& \ x^2 \mathrm{Li}_1(e^{-1/x}) + 5x^3 \mathrm{Li}_2(e^{-1/x}) \nonumber \\
&+12x^4 \mathrm{Li}_3(e^{-1/x})+12x^5 \mathrm{Li}_4(e^{-1/x}),
\end{align} 
with \mbox{$\mathrm{Li}_m(z) = \sum_{l=0}^\infty z^l/l^m$} the m$^{th}$ polylogarithm of $z$. 

Close to $T_{\sf Q}$ one typically expects that $k_{\sf B}T \gg \Delta_1$, and therefore one can make the approximation,
\begin{align}
\frac{1}{T_1^{\sf 1-mag}} \approx
\mathcal{F}_{\sf 1-mag}
\frac{ 6(g\mu_B)^2  \hbar  \gamma_N^2   V_{\sf cell}^2}{\pi^3\bar{v}_s^6} 
(k_B T)^5.
\end{align}


In the context of identifying spin-nematic order, the contribution to the $1/T_1$ relaxation rate from transverse fluctuations is unwanted, since it could potentially mask the signal due to the Goldstone mode fluctuations.
Fortunately, it is possible to suppress the effect of transverse fluctuations by careful choice of nucleus and magnetic field direction.
The form-factor tensor $ \mathcal{F}^{\alpha \beta} \hspace{-0.5mm}({\bf q},{\bf h}_{\sf ext})$ [Eq.~\ref{eq:1/T1}] acts as a filter of spin fluctuations\cite{mila89,smerald11}, and can therefore be used to filter out the low-energy, transverse fluctuations of the partially-polarised moment. 


Denoting the wavevector at which 1-magnon excitations have a minima as ${\bf q}_1$, it is sometimes possible to choose the field direction such that the form factor close to ${\bf q}_1$ is given by,
\begin{align}
\mathcal{F}_{\sf 1-mag}({\bf q}_1+\delta  {\bf q} ,{\bf h}_{\sf ext}) \propto 
\delta {\bf q}^2.
\label{eq:F1mag}
\end{align}  
In such a situation, there is no contribution to the relaxation rate from fluctuations at ${\bf q}_1$, and fluctuations at wavevectors close to ${\bf q}_1$ are suppressed according to $\delta {\bf q}^2$.
In consequence the contribution of 1-magnon processes to the relaxation rate goes as\cite{smerald11},
\begin{align}
\frac{1}{T_1^{\sf 1-mag}} \propto
T^7,
\end{align}
for $k_{\sf b} T \gg \Delta_1$.
In special situations it may be possible to have higher powers of $\delta {\bf q}$ in Eq.~\ref{eq:F1mag}, and therefore even larger suppressions of $1/T_1$.


In Section~\ref{sec:formfactor} we show a worked example of a nuclear environment in which the low-energy, 1-magnon excitations are filtered out by the form factor, but the Goldstone mode fluctuations are not.
The environment we consider is relevant, for example, to P NMR in BaCdVO(PO$_4$)$_2$, and experiments of this type have already been performed in the magnetically-ordered, low-field regime\cite{roy11}. 
We expect that this situation is relatively common, especially when the symmetry of the nuclear environment is high.

\section{Worked example: $^{31}$P NMR in B\lowercase{a}C\lowercase{d}VO(PO$_4$)$_2$}
\label{sec:formfactor}

In Section~\ref{sec:QuadT1} of this Article we have developed a theory of NMR $1/T_1$ 
relaxation rates in a quantum spin-nematic, showing how quadrupolar 
fluctuations lead to characteristic structure in $1/T_1$ at the onset of spin-nematic order 
[cf. Fig. \ref{fig:AFQT1}].
In Section~\ref{sec:transverseT1} we addressed the role of completing dipolar fluctuations, 
and argued that, for an appropriate choice of nucleus and field orientation, these would
largely decouple from $1/T_1$.
In this Section, we develop a worked example of NMR in a material which is a candidate 
for spin-nematic order, and use this to demonstrate how the orientation of magnetic field can 
be used to suppress the contribution of dipolar fluctuations to $1/T_1$.
This analysis closely follows in spirit Ref.~[\onlinecite{smerald11,smerald-thesis}], 
which contain a detailed discussion of the role of form factors in measurements of $1/T_1$.


The material  we consider is the quasi-two dimensional Mott insulator BaCdVO(PO$_4$)$_2$.
%
BaCdVO(PO$_4$)$_2$ \cite{nath08,tsirlin09} is one example of broader family of 
magnetic insulators in which spin-1/2 $V^{4+}$ ions form square-lattice planes.   
This family of materials, which includes the vanadylphosphates 
Pb2VO(PO$_4$)2 \cite{kaul04,shpanchenko06,skoulatos09,nath08}
and SrZnVO(PO$_4$) \cite{skoulatos09,nath08}, 
and the closely-related oxometalates  Li$_2$VOSiO$_4$ 
and Li$_2$VOGeO$_4$ \cite{millet98,melzi00,melzi01,carretta02,rosner02,rosner03,bombardi04},   
have been widely modelled in terms of a spin-1/2 $J_1$-$J_2$ 
Heisenberg model [see Eq.~\ref{eq:HJ1J2}], in which first- and second-neighbour 
interactions on a square lattice compete [see e.g.~\cite{nath08}].
For ferromagnetic first-neighbour interaction $J_1$, the model is known to 
support bond-centered quantum spin-nematic order, with the 
two-sublattice antiferroquadrupolar structure shown in 
Fig.~\ref{fig:SNsquare} [\onlinecite{shannon06}].   
The tendency towards spin-nematic order is particularly strong in applied
magnetic field, approaching saturation \cite{shannon06,sindzingre09,ueda15}.


Published parameters \cite{nath08} place BaCdVO(PO$_4$)$_2$ in a highly-frustrated 
region of the $J_1$-$J_2$ phase diagram, where spin-nematic order competes
with 1-magnon instabilities at ${\bf q} = (\pi,0)$ and ${\bf q} = (0,\pi)$ (see Ref.~[\onlinecite{shannon06,sindzingre09,shindou09,shindou11,shindou13,smerald15,ueda15}]).
This implies that, close to saturation, there will be a gapped spin-wave excitation
at this wave vector, as shown in Fig.~\ref{fig:chiperp}, in addition to excitations of any spin-nematic order.  
Both will contribute to NMR $1/T_1$.

\begin{figure}[t]
\centering
\includegraphics[width=0.49\textwidth]{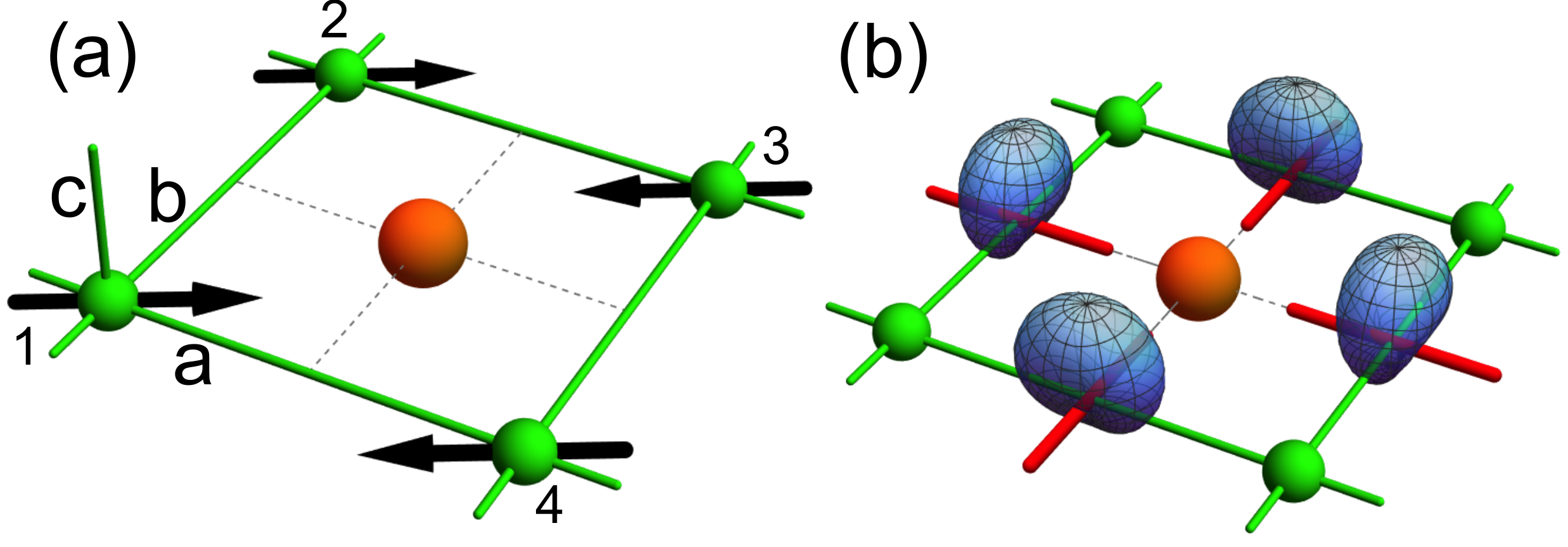}
\caption{\footnotesize{
Local environment of a nuclear site (orange sphere) in a square-lattice, spin-1/2 frustrated ferromagnet.
For example this could be a P nuclear spin in BaCdVO(PO$_4$)$_2$.
$1/T_1$ relaxation occurs due to coupling between the nuclear moment and a square plaquette 
of neighbouring electronic spins (associated with green spheres).
(a) Transverse excitations of the partially-polarised moment at $(\pi,0)$ (shown by black arrows) 
are gapped in the spin-nematic state but can still give a large contribution to the relaxation rate.
By applying the polarising and NMR fields in the c direction, this unwanted contribution can 
be suppressed.
(b) Two-sublattice, bond-centred AFQ order exists in the plane perpendicular to the magnetic field.
This can be parametrised in terms of a set of directors (red) and the blue surface shows the 
probability distribution for the spin moment associated with the bond.
Fluctuations of the quadrupole order parameter result in small spin-dipole flucuations, and 
this can contribute to the $1/T_1$ relaxation rate.
}}
\label{fig:formfactor}
\end{figure}

In what follows, we explore what can be learned about spin-nematic order, of
the type proposed for BaCdVO(PO$_4$)$_2$, from NMR experiments carried 
out on a nucleus at the centre of a square-lattice plaquette, as shown in Fig.~\ref{fig:formfactor}.
To a good approximation, this is the environment of P atoms in 
BaCdVO(PO$_4$)$_2$\cite{nath08,tsirlin09}, and we note that $^{31}$P NMR 
has already been carried out on BaCdVO(PO$_4$)$_2$ at low magnetic 
field\cite{roy11}. 
As discussed in Section~\ref{sec:introduction}, spin-nematic order in BaCdVO(PO$_4$)$_2$ 
cannot lead to splitting of NMR lines, of the type used to diagnose collinear antiferromagnetic order
in Li$_2$VOSiO$_4$ \cite{melzi00,melzi01}.
Nonetheless, fluctuations of the spin-nematic order parameter will contribute 
to the $1/T_1$ measured at the P nuclear site.   
And we will argue that, for a suitable choice of field orientation, the contribution to 
$1/T_1$ from dipolar spin fluctuations at this nuclear site is greatly suppressed, 
revealing the tell-tale fluctuations of spin-nematic order.


We start by assuming that the P nuclear spin interacts via a transferred hyperfine interaction 
with the electronic spins on the four neighbouring magnetic atoms [see Fig.~\ref{fig:formfactor}], in keeping with 
earlier analysis of $^{29}$Si NMR in Li$_2$VOSiO$_4$ \cite{melzi00,melzi01}.  
The internal magnetic field at the P nuclear site is given by,
 \begin{align}
{\bf h}_{\sf int}(t) = \sum_i\boldsymbol{\mathcal{A}}_i . {\bf m}_i(t),
\end{align} 
where $\boldsymbol{\mathcal{A}}_i$ is the nuclear-electron coupling tensor and ${\bf m}_i(t)$ is the 
magnetic moment associated with the  $i$th magnetic ion of the square plaquette. 
The form factor for $1/T_1$ relaxation is given by\cite{smerald11},
\begin{align}
\mathcal{F}^{\alpha\beta}({\bf q},{\bf h}_{\sf ext})
&= \sum_{\gamma,\delta} 
 \left[ R^{x \gamma}_{{\bf h}_{\sf ext}} R^{x \delta}_{{\bf h}_{\sf ext}} + R^{y \gamma}_{{\bf h}_{\sf ext}} R^{y \delta}_{{\bf h}_{\sf ext}} \right] 
\mathcal{A}_{\bf q}^{\gamma \alpha} \mathcal{A}_{-{\bf q}}^{\delta \beta},
\label{eq:formfactorgeneral}
\end{align}
where ${\bf R}_{{\bf h}_{\sf ext}}$ is a rotation matrix relating the direction of the external magnetic 
field to the crystallographic coordinate axes and,
\begin{align}
\mathcal{A}_{\bf q}^{\alpha\beta}=\sum_i e^{i{\bf q}.{\bf r}_i} \mathcal{A}_i^{\alpha\beta}.
\end{align}


The symmetry environment of a square plaquette is $D_{4h}$ (see Fig.~\ref{fig:formfactor}), 
and it follows that the nuclear electron coupling tensor is given by,
\begin{align}
\boldsymbol{\mathcal{A}}_{\bf q}=
4\left(
\begin{array}{ccc}
\mathcal{A}^{\sf aa} c_{\sf a} c_{\sf b} & -\mathcal{A}^{\sf ab} s_{\sf a} s_{\sf b} & 0 \\
-\mathcal{A}^{\sf ab} s_{\sf a} s_{\sf b} &\mathcal{A}^{\sf aa} c_{\sf a} c_{\sf b} & 0 \\
0 & 0 & \mathcal{A}^{\sf cc} c_{\sf a} c_{\sf b}
\end{array}
\right),
\end{align}
where,
\begin{align}
c_{\sf a} &=\cos  \frac{q_{\sf a} }{2}, \quad
c_{\sf b} =\cos  \frac{q_{\sf b} }{2}, \nonumber \\
s_{\sf a} &=\sin  \frac{q_{\sf a} }{2}, \quad
s_{\sf b} =\sin  \frac{q_{\sf b} }{2}.
\end{align} 
For a field applied parallel to the c-axis, the relevant form factor is,
\begin{align}
\mathcal{F}^{\sf cc}({\bf q},h^{\sf c}) & = 16 \left(\mathcal{A}^{\sf cc}\right)^2 
\cos^2 \frac{q_a }{2} \cos^2 \frac{q_b }{2}.
\end{align}    
Close to the wavevector ${\bf q}= (\pi,0)$, this can be approximated by,
\begin{align}
\mathcal{F}^{\sf cc}((\pi+\delta  q_{\sf a},\delta  q_{\sf b}),h^{\sf c}) \approx 
4 \left(\mathcal{A}^{\sf cc}\right)^2 
\delta q_a^2.
\end{align}  
This form factor is zero at ${\bf q} = (\pi,0)$, and therefore fluctuations at this wavevector 
do not contribute to the $1/T_1$ relaxation rate.
Fluctuations close to $(\pi,0)$ do contribute to the NMR relaxation, but with a suppression 
factor of $\delta q_{\sf a}^2$.
An equivalent analysis applies in the vicinity of ${\bf q} = (0,\pi)$.


Form-factor suppression of the 1-magnon excitations is only useful if the form factor remains 
finite at the wavevector associated with the fluctuations of spin-nematic order.
Here we show that this is indeed the case.


The AFQ order parameter lives on the bonds of the square lattice (see Fig.~\ref{fig:SNsquare} 
and Fig.~\ref{fig:formfactor}), and therefore spin fluctuations associated to rotations of the order 
parameter are also bond centred.
The internal magnetic field at the nuclear site can be expressed as,
 \begin{align}
{\bf h}_{\sf int}(t) = \sum_{\langle ij \rangle }\boldsymbol{\mathcal{B}}_{ij} . {\bf m}_{ij}(t),
\end{align} 
where ${\bf m}_{ij}(t)$ is a bond-centred magnetic moment and $\mathcal{B}_{ij}$ is the nuclear-bond coupling tensor.
$\mathcal{B}_{ij}$ can be related to $\mathcal{A}_i$ according to $\mathcal{B}_{ij} = (\mathcal{A}_i+\mathcal{A}_j)/2$.
In consequence one finds,
\begin{align}
\boldsymbol{\mathcal{B}}_{\bf q}=
4\left(
\begin{array}{ccc}
\mathcal{A}^{\sf aa} c_{\sf a} c_{\sf b} & 0 & 0 \\
0 &\mathcal{A}^{\sf aa} c_{\sf a} c_{\sf b} & 0 \\
0 & 0 & \mathcal{A}^{\sf cc} c_{\sf a} c_{\sf b}
\end{array}
\right).
\end{align}


For field applied in the c direction the spin-nematic directors lie in the ab plane and spin fluctuations are 
parallel to the c direction and at ${\bf q}\approx 0$.
The relevant form factor is therefore,
\begin{align}
\mathcal{F}^{\sf cc}((0,0),h^{\sf c}) = 
16 \left(\mathcal{A}^{\sf cc}\right)^2.
\end{align}  
This is finite, and therefore there is no form factor suppression of the Goldstone mode contribution 
to $1/T_1$.


It follows that $^{31}$P NMR in BaCdVO(PO$_4$)$_2$ {\it is} sensitive to spin-nematic order.
Following the analysis in Section~\ref{sec:QuadT1}, the relevant contribution to $1/T_1$ 
should be particularly marked near the transition from the polarised paramagnet 
into the spin-nematic state, where it leads to a sharp cusp, and step-like jump in $1/T_1$, 
as illustrated in Fig.~\ref{fig:AFQT1}.
 

This analysis demonstrates that it is possible to use the orientation of magnetic field to select
between quadrupolar and dipolar fluctuations in measurements of $1/T_1$.
And, while we have considered the concrete example of $^{31}$P NMR in BaCdVO(PO$_4$)$_2$, 
the underlying principle of using a form factor to select between different fluctuations is 
much more general.
Exactly how this filtering effect works depends on the nuclear site chosen.
A good rule of thumb is that, the higher the symmetry of the nuclear environment, 
the more likely it will be that a field-direction can be found for which fluctuations 
of spin-nematic order come to the fore.

\section{1/T$_1$ relaxation at low temperature}
\label{sec:lowT}

For completeness we have also calculated the spin-lattice relaxation rate at low temperatures 
(see Appendix~\ref{app:lowT}).
We consider the case $T \ll \Delta_1$, where the contribution from 1-magnon 
excitations is exponentially suppressed, and, therefore, only the Goldstone mode fluctuations 
contribute significantly to $1/T_1$.


The low-temperature relaxation rate in the spin nematic is controlled by the 
long-wavelength excitations of the quadrupolar order parameter.
These can be described by a sigma-model model, as detailed in Ref.~[\onlinecite{smerald13,smerald-thesis,smerald15}].
We restrict our focus to the Goldstone mode excitation characteristic of the AFQ state.


Spin-dipole fluctuations are associated with time derivatives of the field, and, in 3D and for 
$T \ll T_N$ this leads to (see Appendix~\ref{app:lowT}),   
\begin{align}
\frac{1}{T_1^{\sf AFQ}}  &\approx  \mathcal{F} \frac{9\tilde{I}_{\sf hT} \hbar V_{\sf cell}^3}{256\pi^5} 
   (m_{\sf s}^{\sf eff})^2 
       \gamma_N^2   
    \frac{\left(k_{\sf B}T \right)^7}{\chi_{h}^{\sf Q,z} v^9},   
\end{align} 
where $\tilde{I}_{hT}  \approx 146.5$, 
$(m_{\sf s}^{\sf eff})^2 = \sqrt{3}(g\mu_{\sf B})^2 \left| \langle {\bf Q}_{\perp} \rangle \right|/2$, 
$\chi_{h}^{\sf Q,z}$ is the susceptibility associated with the Goldstone mode, 
and $v$ is its velocity.
\footnote{We note that in Ref.~[\onlinecite{barzykin91}] $1/T_1 \propto1/T^7$ and 
$1/T_1 \propto1/T^{11}$ power laws were quoted without explanation.}


We note that relaxation due to excitation of acoustic phonons also has a $1/T_1 
\propto T^7$ behaviour\cite{abragam-book}, but could in principle be distinguished through its
(lack of) magnetic field dependence.
In practice, observation of this type of power-law behaviour is likely to be very challenging, because of the slow relaxation.

\section{Discussion and Conclusion}
\label{sec:conclusion}

In this article we have studied the behaviour of the $1/T_1$ relaxation rate in a quantum spin-nematic state close to saturation.
We have concentrated on extracting the universal features, and the main result is shown in Fig.~\ref{fig:AFQT1}.
We find that the most distinctive feature occurs at the thermal phase transition between the spin-nematic state and the partially-polarised paramagnet, where $1/T_1$ shows a step-like feature and a sharp cusp, but no critical divergence.


The underlying physical mechanism for relaxation is a small spin-dipole fluctuation driven by the dynamics of the antiferroquadrupolar (AFQ) order parameter.
This creates a fluctuating internal magnetic field at the nuclear site, which couples to the nuclear spin lattice.
Since the spin-dipole fluctuations are suppressed on approaching the AFQ ordering vector, they do not experience critical slowing down, and therefore there is no divergence in $1/T_1$ at the critical point.


There are two obvious questions that arise:
\begin{enumerate}
\item How feasible is it to measure a $1/T_1$ relaxation rate with the form shown in Fig.~\ref{fig:AFQT1}? 
\item If measurement proves possible, how characteristic are these features of the spin-nematic state?
\end{enumerate}


To answer the first question we have studied the $1/T_1$ relaxation rate arising from transverse excitations of the partially-polarised moment.
In spin-1/2, frustrated ferromagnets close to saturation, these 1-magnon excitations often have a gap that is small compared to the ordering temperature, $\Delta_1 \ll T_{\sf N}$.
As a result there is a contribution to the relaxation rate that goes as  $1/T_1 \propto T^5$.
From our approach, it is not possible to determine in a quantitative way how this compares to the contribution to $1/T_1$ from fluctuations of the AFQ Goldstone mode, since there are a number of phenomenological parameters.
However, knowing the power-law that governs this type of relaxation should make it possible to fit the 1-magnon contribution and subtract it from the experimental data.
Furthermore, we show that by careful choice of nuclear site and magnetic-field direction, it is possible to suppress the 1-magnon contribution by using the form factor as a filter.
In consequence the 1-magnon relaxation rate goes as $1/T_1 \propto T^7$, and in special circumstances it may be possible to suppress it even further.


To address the second question, we suggest  that measurement of the step-like feature in the $1/T_1$ relaxation rate shown in Fig.~\ref{fig:AFQT1} would be strongly suggestive of spin-nematic order, especially when taken in conjunction with other information, and we set out the reasons for this below.
However, unlike inelastic neutron scattering, where it is in principle possible to measure the Goldstone mode associated with AFQ order\cite{smerald15}, $1/T_1$ does not provide direct evidence for the spin-nematic state.


Firstly, one may imagine that a similar step-like feature could appear in a magnetically ordered system with an anisotropy driven gap.
However, the absence of a critical divergence at any temperature is a clear indication that there is no magnetic long-range order.
This could also be confirmed by the absence of splitting of spectral lines in static NMR measurements.


Secondly, one could imagine a similar step-like feature appearing in the saturated paramagnet, where it would be associated with the gap to 1-magnon excitations.
One way to definitively rule out this possibility would be to study the evolution of this step-like feature with magnetic field.
For a saturated paramagnet, the size of the gap increases with increasing magnetic field, and therefore the step-like feature would move to higher temperature.
For the spin-nematic state, the critical temperature is expected to be either constant or decreasing with increasing magnetic field (see Fig.~\ref{fig:phasediag}), and therefore the step-like feature would move to lower temperature.


After ruling out these two possibilities, one can conclude that there is a non-trivial 
state 
without spin-dipole order.
In the absence of other information, options include a valence-bond solid state, 
with a $1/T_1$ step at the temperature at which singlet to triplet excitations become 
energetically possible, or a gapped spin liquid, with the $1/T_1$ step at a temperature 
comparable to the gap.
However, it will in general be possible to rule out these possibilities from other 
considerations.
For example, if there is good evidence that the first-neighbour coupling, $J_1$,
is ferromagnetic, this would be incompatible with a valence-bond solid state.


In conclusion, NMR $1/T_1$ measurements resembling Fig.~\ref{fig:AFQT1}, 
combined with other considerations, would provide strong evidence for the existence 
of a spin-nematic state.
We end with the hope that this analysis can help to motivate further experiments 
on the many candidate materials for spin-nematic order, and that this ghostly 
transition will, in time, be revealed by its spectral features. 


\section*{Acknowledgments}   
The authors are pleased to acknowledge enlightening discussions with
Pietro Carretta, 
Tsutomu Momoi, 
Kazuhiro Nawa, 
Karlo Penc, 
Masashi Takigawa
and Hiroaki Ueda.   
This work was supported by the Theory of Quantum Matter Unit of the Okinawa Institute
of Science and Technology Graduate University, by EPSRC Fellowship 
and by EPSRC grants EP/C539974/1 and EP/G031460/1

\appendix

\section{Theory of NMR relaxation rates in a conventional canted antiferromagnet}
\label{app:NMR-AFM}

For comparative purposes, we develop a critical theory of the relaxation rate in a conventional 
canted antiferromagnet (AFM), where $1/T_1$ is known to diverge for $T \to T_N$~[\onlinecite{moriya62,borsa-book,ziolo88, engelsberg80}].


Spin excitations in the AFM can be described phenomenologically using time-dependent Ginzburg-Landau 
theory\cite{chaikin-book,ma-book,halperin69a,halperin69b,hohenberg77}, written in terms of the order 
parameter ${\bf m}_{\sf s}({\bf r})$, and its associated canting field $ l^{\sf z}({\bf r})$,
\begin{align}
\mathcal{H}_{\sf AFM} \hspace{-0.7mm} \approx  \hspace{-0.7mm}
\int \hspace{-0.7mm} d^3r \left[ \frac{\alpha\tilde{t}}{2}{\bf m_{\sf s}}^2+ \frac{K}{2} (\nabla {\bf m}_{\sf s})^2  
+ \frac{u}{4} {\bf m_{\sf s}}^4 +\frac{1}{2\chi_\perp}(l^{\sf z})^2 \right]
\label{eq:GL-AFM}
\end{align}
where a term $hl^{\sf z}$ is ignored as irrelevant.
Here \mbox{$\tilde{t} = (T-T_{\sf N})/T_{\sf N}$}, $K$ is a generalised elastic constant, 
$\chi_\perp$ the transverse susceptibility, $\alpha > 0$ and $u > 0$ are positive constants 
and magnetic field is applied parallel to the ${\sf z}$-axis.
Length scales are measured in units of the lattice spacing and energy 
scales in units of $k_{\sf B}T_{\sf N}$.
Expanding, 
$${\bf m}_{\sf s}({\bf r})=\eta({\bf r}) \hat{\bf n}({\bf r})=[\langle{\eta}\rangle+\delta \eta({\bf r})] \hat{\bf n}({\bf r}),$$ 
where 
$\hat{\bf n}({\bf r})$ is a 2D unit vector, results in,
\begin{align}
\mathcal{H}_{\sf AFM} \approx 
\int  d^3r & \left[\frac{\alpha\tilde{t}}{2} \eta^2+ \frac{K}{2} (\nabla \eta)^2  
+ \frac{u}{4} \eta ^4  \right. \nonumber \\
&\left. +\frac{\rho_{\sf s}(\eta)}{2} (\nabla \hat{\bf n})^2
+ \frac{(l^{\sf z})^2}{2\chi_\perp} \right],
\label{eq:GL-AFM-eta}
\end{align}
where the spin stiffness is defined by \mbox{$\rho_{\sf s}(\eta)=K\eta^2$}.
The Ginzburg-Landau theory predicts a diverging correlation length \mbox{$\xi \propto \tilde{t}^{-\nu}$} 
approaching the critical point, with mean-field exponent $\nu=1/2$.


The critical divergence of $\xi$ is accompanied by a critical slowing down in fluctuations of AFM order.
%
This can be described  by adding dynamics to the static Ginzburg-Landau model, to capture the 
way in which the slow, long-wavelength fluctuations of AFM order are damped by interaction 
with short-wavelength fluctuations.  
Following Ref.'s~[\onlinecite{chaikin-book,ma-book,halperin69a,halperin69b,hohenberg77}], we write,
\begin{align}
 \partial_t \delta\eta({\bf r},t) = -\Gamma K (\xi^{-2} -\nabla^2 ) \delta\eta({\bf r},t) +\zeta({\bf r},t),
 \label{eq:eta-EOM}
\end{align}
where time is measured in units of $\hbar/k_{\sf B}T_{\sf N}$, $\Gamma^{-1}$ sets the rate of damping, 
$u\rightarrow 0$, and $\zeta({\bf r},t)$ is a (white) noise term with correlation function,
\begin{align}
\langle \zeta({\bf r},t) \zeta({\bf r}^\prime,t^\prime) \rangle = 2k_{\sf B}T\Gamma \delta({\bf r}-{\bf r}^\prime) \delta(t-t^\prime).
\label{eq:Gamma-noise}
\end{align}
It follows that the correlation time is given by \mbox{$\tau_\eta({\bf q})\approx\Gamma^{-1}\chi_{\bf q}$}, 
with \mbox{$\chi_{\bf q}=K^{-1}(\xi^{-2}+ {\bf q}^2)^{-1}$}.
Approaching $T_{\sf N}$,  $\tau_\eta(0)$ diverges as $\tau_\eta(0) \propto \xi^z$ with 
dynamical exponent $z=2$.


In the case of the antiferromagnet, fluctuations of the field $\eta$ are directly related to fluctuations 
of the spin-dipole moments, since, for moments canted in the $x$-$z$ plane, $S^{\sf x}({\bf r},t)\propto \pm \eta({\bf r},t)$.
Therefore, the dynamic susceptibility associated with $\eta$ controls the critical behaviour of the $1/T_1$ relaxation rate.
The fluctuation-dissipation theorem allows the imaginary part of the dynamic susceptibility to be calculated 
from Eq.~\ref{eq:eta-EOM} and Eq.~\ref{eq:Gamma-noise} as,
 \begin{align}
\frac{k_{\sf B}T}{\hbar\omega} \Im m\{\chi^{\sf zz}_\eta({\bf q},\omega)\} 
\approx (g_{\sf l}\mu_{\sf B})^2 \frac{k_{\sf B}T}{(k_{\sf B}T_{\sf N})^2} \frac{ \Gamma}{ \Gamma^2 \chi_{\bf q}^{-2}+\omega^2},
\label{eq:AFM-chi}
 \end{align}
 where $g_{\sf l}$ is the Land\'{e} g-factor.
Thus the dynamic susceptibility diverges approaching the critical point for both 
${\bf q}\rightarrow 0$ and $\omega \rightarrow 0$.


The mean-field behaviour of the $1/T_1$ relaxation rate can be calculated from Eq.~\ref{eq:1/T1}  
and Eq.~\ref{eq:AFM-chi} as,
\begin{align}
\frac{1}{T_1^{\sf AFM}} \approx  \frac{(g_{\sf l}\mu_{\sf B})^2 }{4\pi^2}\gamma_{\sf N}^2 \mathcal{F} 
\frac{k_{\sf B}T}{(k_{\sf B}T_{\sf N})^2} \frac{1}{\Gamma K^2}f(\Lambda \xi) \xi, 
\label{eq:AFM-1/T1}
\end{align}
where a constant form factor, \mbox{$\mathcal{F}^{\alpha\beta}({\bf q},{\bf h}_{\sf ext}) 
\rightarrow \mathcal{F}$}, is assumed, $\Lambda$ is a momentum cut-off and \mbox{$f(x)=[ \arctan x -x/(1+x^2)]/2$},
where $f(\infty)=\pi/4$.
Approaching the critical point Eq.~\ref{eq:AFM-1/T1} predicts $1/T_1 \propto \xi \propto \tilde{t}^{-\nu}$, i.e. the 
NMR $1/T_1$ relaxation rate diverges with mean-field exponent $\nu=1/2$.
More generally this can be written as\cite{ziolo88},
 \begin{align}
\frac{1}{T_1} \propto \xi^{z+2-d} \propto \tilde{t}^{(d-2-z)\nu},
 \end{align}
where the exponents $z$ and $\nu$ exponents take on values appropriate for an $O(2)$ phase transition.
However experiment confirms that mean-field theory is qualitatively correct in 
predicting a divergence in $1/T_1$ [\onlinecite{borsa-book,ziolo88,engelsberg80}].

\section{NMR 1/T$_1$ at low temperature in the spin nematic state}
\label{app:lowT}

Low temperature measurement of the NMR $1/T_1$ relaxation rate could also be used to identify the spin-nematic state.
Here we show that at low temperature the relaxation rate follows a $1/T_1 \propto T^7$ power-law behavior.


In a conventional collinear antiferromagnet at low temperature\cite{smerald11},
 \begin{align}
\frac{1}{T_1} \approx 
\mathcal{F} \frac{ \hbar V_{\sf cell}^2 m_{\sf s}^2  \gamma_N^2  \Delta^3}{8\pi^3 \bar{v}_s^6 \chi_\perp^2} \
 \Phi  \left( \frac{k_{\sf B} T}{\Delta} \right),
\end{align} 
where a constant form factor $\mathcal{F}$ is assumed, $m_{\sf s}$ is the ordered moment, $V_{\sf cell}$ is the volume of a unit cell, $\Delta$ is an energy gap, $\bar{v}_s$ is the geometric mean of the spin wave velocities, $\chi_\perp$ is the perpendicular susceptibility and,
\begin{eqnarray} 
\Phi (x) = x^2 \mathrm{Li}_1(e^{-1/x}) + x^3 \mathrm{Li}_2(e^{-1/x}),
\end{eqnarray} 
with
\mbox{$\mathrm{Li}_m(z) = \sum_{l=0}^\infty z^l/l^m$} the m$^{th}$ polylogarithm of $z$. 
For temperatures $T\gg \Delta$,
 \begin{align}
\frac{1}{T_1} \approx 
\mathcal{F} \frac{ \hbar  V_{\sf cell}^2}{48\pi } m_{\sf s}^2  \gamma_N^2 
 \frac{(k_{\sf B}T)^3}{\chi_\perp^2 \bar{v}_s^6 },
\end{align} 
while for temperatures $T\ll \Delta$,
 \begin{align}
\frac{1}{T_1} \approx 
\mathcal{F} \frac{ \hbar  V_{\sf cell}^2}{8\pi^3  }  m_{\sf s}^2  \gamma_N^2 \
\frac{(k_{\sf B}T)^2 \Delta}{\chi_\perp^2 \bar{v}_s^6} e^{-\frac{\Delta}{k_{\sf B}T}}.
\end{align} 
For a discussion of cases where the form factor is not constant see Ref.~[\onlinecite{smerald11}].


In a 2-sublattice AFQ state in large applied magnetic field, $1/T_1$ can be calculated by an analogous method.
The excitations of the quadrupolar order parameter can be described by a non-linear sigma model field theory\cite{smerald13,smerald-thesis,smerald15}, and this can be linearised to give the effective action,
 \begin{align}
&\mathcal{S}_{\sf 2SL} [\phi,h] \approx \frac{1}{2V_{\sf cell}}\int_0^\beta d\tau \int d^3r
\nonumber \\
 \{ 
&\chi_{h}^{\sf Q,z}  (\partial_\tau \phi)^2 
+\sum_\mu \rho_{h}^{\sf Q,z} (\partial_\mu \phi)^2 
+ \chi_{h}^{\sf Q,z}  \Delta_{\sf ex}^2 \phi^2 \},
\label{eq:AFQ-action}
\end{align}
where $\rho_{h}^{\sf Q,z}$ is the order parameter stiffness, $\chi_{h}^{\sf Q,z}$ the susceptibility and $\Delta_{\sf ex}$ a very small energy gap due to anisotropy.
This action describes an almost gapless mode with dispersion of the form $\omega_{\bf k}=\sqrt{\Delta_{\sf ex}^2+v^2k^2}$, where $v=\sqrt{\rho_{h}^{\sf Q,z}/\chi_{h}^{\sf Q,z}}$.


\begin{figure}[h]
\centering
\includegraphics[width=0.45\textwidth]{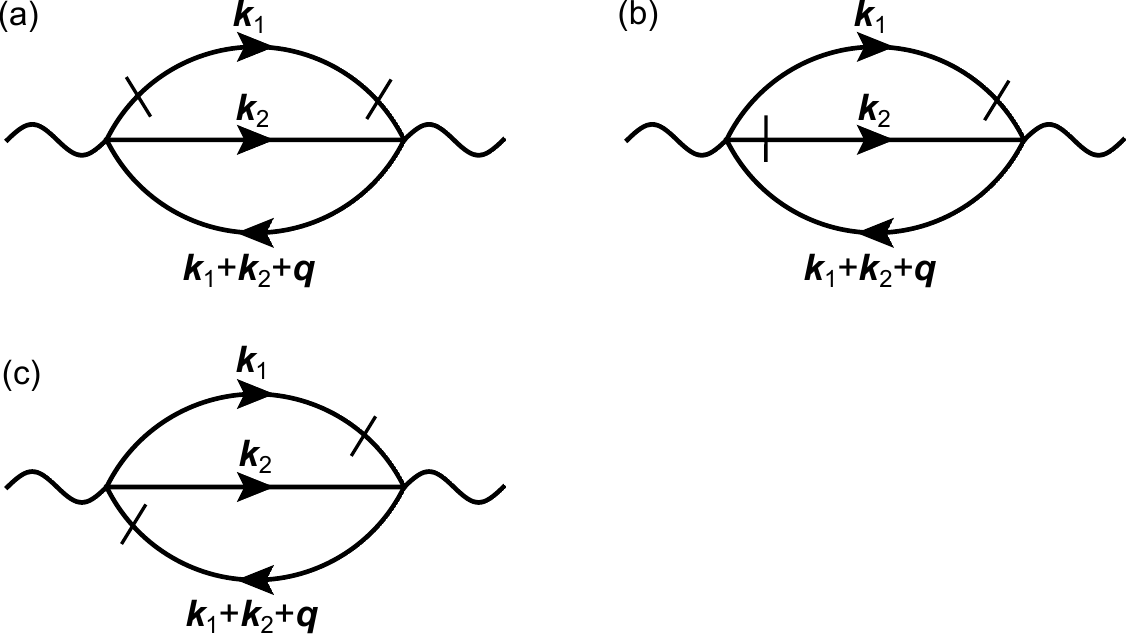}
\caption{\footnotesize{
Diagrams used to calculate the low-temperature, dynamic spin susceptibility in the spin-nematic state.
Slashes denote differentiation of the $\phi$ field with respect to time.
It follows that $1/T_1\propto T^7$.
}}
\label{fig:lowT-diagrams}
\end{figure}


The action can be used to calculate the dynamic spin susceptibility, and $1/T_1$ follows from Eq.~(\ref{eq:1/T1}) in the main text.
The lowest order contribution to the relaxation rate is the `three magnon' term, which involves averages over six $\phi$ fields with two time derivatives. 
Evaluating the diagrams shown in Fig.~\ref{fig:lowT-diagrams} gives,
\begin{align}
\frac{1}{T_1} &\approx  \mathcal{F} \frac{9\hbar V_{\sf cell}^3 }{32} (m_{\sf s}^{\sf eff})^2 \gamma_N^2  \frac{I(T)}{\chi_{h}^{\sf Q,z}}  ,
\end{align} 
where \mbox{$(m_{\sf s}^{\sf eff})^2 = \sqrt{3}(g\mu_{\sf B})^2 \left| \langle {\bf Q}_{\perp} \rangle \right|/2$}, a constant form factor $\mathcal{F}$ has been assumed and,
\begin{align}
&I(T)= 
 \frac{\Delta_{\sf ex}^7 }{8\pi^5v^9}  \left(\frac{k_{\sf B}T}{\Delta_{\sf ex}} \right)^7  \int_{\frac{\Delta_{\sf ex}}{k_{\sf B}T} }^\infty dx_1    dx_2  
\sqrt{ x_1^2-\left(\frac{\Delta_{\sf ex}}{k_{\sf B}T} \right)^2} \nonumber \\
& \sqrt{ x_2^2-\left(\frac{\Delta_{\sf ex}}{k_{\sf B}T} \right)^2 }
 \sqrt{ (x_1+x_2)^2-\left(\frac{\Delta_{\sf ex}}{k_{\sf B}T} \right)^2} \nonumber  \\
& \times \left( x_1^2+ x_1 x_2 + x_2^2 \right)
 \frac{e^{x_1+x_2} }{( e^{x_1}-1)( e^{x_2}-1) ( e^{x_1+x_2}-1)}.
 \label{eq:T^7integral}
\end{align} 
At temperatures \mbox{$T\ll \Delta_{\sf ex}$} this can be approximated by,
\begin{align}
\frac{1}{T_1} &\approx \mathcal{F} \frac{27\sqrt{3}\tilde{I}_{\sf lT} \hbar  V_{\sf cell}^3}{128\pi^5}  (m_{\sf s}^{\sf eff})^2 \gamma_N^2  
 \frac{\Delta_{\sf ex}^4 (k_{\sf B}T)^3}{\chi_{h}^{\sf Q,z} v^9}  
 e^{-\frac{2\Delta_{\sf ex}}{k_{\sf B}T}},
\end{align} 
where,
\begin{align}
\tilde{I}_{\sf lT} =  \int_0^\infty dy_1    dy_2 \sqrt{y_1y_2} e^{-(y_1+y_2)} \approx 0.79.
 \end{align} 
For temperatures $T\gg \Delta_{\sf ex}$,
\begin{align}
\frac{1}{T_1} &\approx  \mathcal{F} \frac{9\tilde{I}_{\sf hT} \hbar V_{\sf cell}^3}{256\pi^5} 
   (m_{\sf s}^{\sf eff})^2 
       \gamma_N^2   
    \frac{\left(k_{\sf B}T \right)^7}{\chi_{h}^{\sf Q,z} v^9},   
\end{align} 
where,
\begin{align}
\tilde{I}_{\sf hT}=\int_0^\infty &dx_1    dx_2 \ x_1x_2(x_1+x_2)  
 \left( x_1^2+ x_1 x_2 + x_2^2 \right) \nonumber \\
&\times \frac{e^{x_1+x_2} }{( e^{x_1}-1)( e^{x_2}-1) ( e^{x_1+x_2}-1)} \approx 146.5.
\end{align} 

\bibliographystyle{apsrev4-1}
\bibliography{bibfile}

\end{document}